**Cell-cycle regulated transcription associates with DNA replication timing in yeast and human**

Hunter B. Fraser

Department of Biology, Stanford University, Stanford CA 94305.  Email: hbfraser@stanford.edu.

**Abstract**

**Background:** Eukaryotic DNA replication follows a specific temporal program, with some genomic regions consistently replicating earlier than others, yet what determines this program is largely unknown. Highly transcribed regions have been observed to replicate in early S-phase in all plant and animal species studied to date, but this relationship is thought to be absent from both budding yeast and fission yeast. No association between cell-cycle regulated transcription and replication timing has been reported for any species.

**Results:** Here I show that in budding yeast, fission yeast, and human, the genes most highly transcribed during S-phase replicate early, whereas those repressed in S-phase replicate late. Transcription during other cell-cycle phases shows either the opposite correlation with replication timing, or no relation. The relationship is strongest near late-firing origins of replication, which is not consistent with a previously proposed model—that replication timing may affect transcription—and instead suggests a potential mechanism involving the recruitment of limiting replication initiation factors during S-phase.

**Conclusions:** These results suggest that S-phase transcription may be an important determinant of DNA replication timing across eukaryotes, which may explain the well-established association between transcription and replication timing.

**Background**

The timing of DNA replication during S-phase of the cell cycle plays an important role in genome integrity, the mutational spectrum, and a wide range of human diseases [1]. Despite many recent advances in our ability to measure the time of replication ($T_{rep}$) across entire genomes [2-7], our understanding of what regulates this timing remains far from complete [1, 8-11]. The time at which origins of replication (ORIs) fire is thought to be determined in M-phase [12] or G1 [13-14], at which point factors such as Cdc45 and Sld3 bind to ORIs that will fire early in the following S-phase [15-16]. These and several other proteins critical for replication initiation are present at copy-numbers lower than the number of ORIs [17-19], and their over-expression advances $T_{rep}$ for many late-firing ORIs in both budding and fission yeast [12,17-20], suggesting that their re-use may be a key step in regulating ORI firing time. However what determines the relative affinities of different ORIs for these limiting factors—and hence their temporal order of initiation—is largely unknown [19].

Among the strongest correlates (and potential determinants) of $T_{rep}$ in metazoans are transcriptional activity and chromatin state. Although transcriptionally active euchromatin has been known to replicate earlier than repressive heterochromatin for over 50 years [11,21], the reason—and even the direction of causation—has remained elusive. The two major models [8,11], not mutually exclusive, are that 1) The euchromatic chromatin structure is more permissive both to transcription and to DNA replication initiation; or 2) $T_{rep}$ itself affects chromatin structure and transcription, as a result of changes in the nuclear milieu during S-phase. The former is most directly supported by experiments altering ORI firing time via manipulation of histone modifications [8-10,18,22-24], whereas the latter is supported by differences in chromatin and transcription of DNA templates injected into cells during either early or late S-phase [8-9,25-26].

Measuring $T_{rep}$ genome-wide in the budding yeast *Saccharomyces cerevisiae* (*Sc*), Raghuraman et al. [2] reported a surprising lack of association between transcription and $T_{rep}$ (with the exception of the eight histone genes, which are highly transcribed in S phase and are replicated early). However this analysis only involved clusters of co-expressed genes, and did not actually compare the highest vs. lowest-expressed genes. Nevertheless, it has been widely interpreted in the literature as indicating the absence of any association, and many authors have speculated as to why budding yeast lacks this relationship [5,8-11]. Similarly the fission yeast *Schizosaccharomyces pombe* (*Sp*) is thought to lack any association between transcription and replication timing [8], though again no systematic comparison has been reported.

**Results**

Because DNA replication is confined to a specific period during the cell cycle, I reasoned that the relationship between $T_{rep}$ and transcription may depend on when in the cell cycle transcription is occurring. The transcription of most genes does not vary greatly throughout the cell cycle, so cannot be used to determine phase-dependent effects. However several hundred genes have been identified in both *Sc* and *Sp* that do vary consistently during the cell cycle [27-28]. I compared the expression levels of these cell cycle-regulated genes measured in synchronized cells [27-28] with the $T_{rep}$ for each gene, to determine if any relationship exists. For both *Sc* and *Sp* expression levels measured in G2 phase, higher expression associated with earlier $T_{rep}$ (Figure 1a). However at other points in the cell cycle the relationship was quite different; mostly notably in M/G1 (*Sc*) or G1 (*Sp*), the relationship reversed, such that highly expressed genes were replicated late (Figure 1a).

To more systematically visualize these patterns, I calculated the correlation between the expression levels of all cell cycle-regulated genes measured in synchronized cultures [27-28] with their $T_{rep}$, separately for each expression data time-point (see Materials and Methods). Plotting these correlation coefficients as a function of the time at which the expression data were sampled, I found a striking relationship: both the strength and direction of the correlation oscillate as a function of cell-cycle stage (Figure 1b). In these plots, positive *r* values represent time-points at which up-regulated genes tend to be replicated late in S phase; negative *r* values indicate times when up-regulated genes are replicated early. Consistent with the results in Figure 1a, in both species of yeast, genes highly expressed in G2 phase are replicated early, while those expressed in late M/G1 are replicated late. The oscillation is observed regardless of the method used to achieve cell-cycle synchronization (Figures S1-S2).

To further characterize this relationship, I plotted a moving average of $T_{rep}$ for the cell cycle-regulated genes in each species, ordered by their time of maximal expression. If expression in certain cell-cycle phases correlates with early or late replication, this should be reflected by troughs or peaks in such a plot. Again in both species a similar trend emerged: $T_{rep}$ reaches a maximum for genes expressed in G1, and a minimum for those expressed in G2 (Figure 1c, Figure S3), consistent with the correlation analysis (Figure 1b). The strong conservation of this pattern was surprising, considering how much the regulation of DNA replication has diverged in the hundreds of millions of years separating these two yeast lineages [29].

Although the strongest association between high mRNA levels and early replication was observed for G2-phase expression levels, it is important to note that this does not imply these genes are maximally *transcribed* in G2. Rather, one would expect maximal transcription to occur in the time leading up to the maximal transcript level, i.e. in S phase. Indeed, plotting mRNA levels for G2-upregulated genes (those with early $T_{rep}$ in Figure 1c), it is clear that their transcript levels show the greatest increase—likely reflecting active transcription—in S phase (Figure S4a). Likewise, genes with late $T_{rep}$ show the opposite pattern: maximal decrease in mRNA levels during S phase (Figure S4b).

The oscillating relationships shown in Figure 1 do not establish whether $T_{rep}$ is more directly associated with transcription in S phase, or in M phase. For example, if M-phase repression led to early $T_{rep}$, S phase induction could be associated with early $T_{rep}$ simply as an indirect consequence, because genes repressed in M phase are typically induced in S phase (Figure S4a). To disentangle the effects of S and M phases, I examined the $T_{rep}$ of genes that are expressed at similar levels throughout the cell cycle. If M-phase repression leads to early $T_{rep}$, then genes repressed throughout the cell cycle would be expected to have early $T_{rep}$, as a result of their repression in M-phase (in this scenario, their S-phase expression levels are not relevant). However if the association is instead due to S-phase induction, genes with constitutive high expression would have earlier $T_{rep}$, because of their active transcription in S phase (in which case M-phase expression levels would be irrelevant). This analysis showed a clear trend: highly expressed genes replicate 5.9 min earlier in *Sc* and 3.0 min earlier in *Sp* (Figure 2). Therefore the results shown in Figure 1 can be entirely, and most parsimoniously, explained by the association of $T_{rep}$ with S-phase transcription; the M-phase relationship is likely to be an indirect side-effect of this. This result also suggests a more general association between transcription and $T_{rep}$ in yeast that extends beyond cell cycle-regulated genes.

To further investigate the connection between S-phase transcription and $T_{rep}$, I tested whether the relationship differed for genes replicated during early vs. late S-phase. In this analysis I separated all cell cycle-regulated genes into ten bins (i.e. deciles) by their $T_{rep}$, and plotted the median G2-phase transcript level (the time point most closely reflecting S-phase transcription; Figure S4a) for each. Across all ten $T_{rep}$ bins, I observed the expected relationship: decreasing expression of bins with increasing $T_{rep}$ (Figure 3a). However closer examination revealed that for both yeast species, this pattern was almost entirely driven by late-replicating genes. In other words, there was no correlation between expression levels and $T_{rep}$ for genes in the first five bins (replicated in early S phase), while in late S phase the relationship was quite strong (Figure 3a). Consistent with this, applying the correlation analysis from Figure 1a to just early or late-replicating genes revealed that the oscillation is entirely driven by replication in late S-phase; genes with early $T_{rep}$ showed no oscillation, and only a weak correlation at nearly all time points (Figure 3b). These results parallel the finding that in mouse, genes replicated in the second half of S-phase show the strongest association between transcript levels and $T_{rep}$ [7].

Another factor that may influence the relationship between S-phase transcription and replication timing is a gene's distance from the nearest ORI. Under the model where chromatin affects both transcription and $T_{rep}$, the strongest association would be expected for genes near ORIs, whereas if instead $T_{rep}$ affects a gene's level of S-phase transcription, the relationship should be independent of distance to the nearest ORI [3]. Separating genes into two classes, ORI-proximal or ORI-distal, the ORI-proximal class showed far stronger oscillations (Figure 3c) (ORI distance cutoffs, chosen to result in approximately equal-sized lists, were 5 kb from the nearest ORI in *Sc* and 10 kb in *Sp*, due to the higher density of known ORIs in *Sc*; results from equal distance cutoffs are shown in Figure S5). Because ORI-proximal genes tend to be replicated earlier than ORI-distal genes, this result could not be an indirect effect of the stronger association for late-$T_{rep}$ genes, as it acts in the opposite direction. This result suggests that the relationship is unlikely to be caused by an effect of $T_{rep}$ on S-phase transcription, which is one of the two major classes of models that have been proposed to explain the transcription/$T_{rep}$ association [8,25-26].

To test whether the relationship between S-phase transcription and replication timing is conserved outside of fungi, I applied the same correlation analysis to cell-cycle gene expression and $T_{rep}$ data from human HeLa cells [6,30]. Analyzing all known HeLa cell-cycle regulated genes [30], I found no significant relationship of any kind (Figure 4a). However applying each of the two filters identified from yeast—late $T_{rep}$ and ORI proximity (within 10 kb)—resulted in clear and significant oscillations, of a magnitude similar to that observed for both yeast species (Figure 4b-c). As observed for yeast, the minimum correlation (indicating early $T_{rep}$ of up-regulated genes) occurred in G2, and the maximum in late M/G1. The fact that the same oscillating relationship exists in human, and that its strength is influenced by the same two factors, suggests that it is likely to be caused by a mechanism conserved between fungi and metazoans.

To put into perspective the strength of the relationship between $T_{rep}$ and cell cycle-regulated gene expression in human, I compared it to the well-established association between $T_{rep}$ and average (asynchronous) gene expression. The latter provides a useful benchmark because it is regarded as a strong relationship that has been observed in numerous studies across diverse metazoans [5-6,8-9]. To facilitate a direct comparison with the results in Figure 4, I used the same $T_{rep}$ data [6] for the same genes, but replaced the cell-cycle synchronized gene expression data [30] with high-coverage RNA-seq data from asynchronous HeLa cells [31]. The correlation between asynchronous expression and $T_{rep}$ was $r = -0.16$ for late $T_{rep}$ genes (the genes represented by the red line in Figure 4b) and $r = -0.15$ for ORI-proximal genes (represented by the blue line in Figure 4c). In both cases, the asynchronous data explained less than a third of the variance in $T_{rep}$ that is explained by S-phase transcription (see Materials and Methods). Differing quality of the two gene expression data sets [30-31] could contribute to this difference; however because RNA-seq is of far higher precision than spotted cDNA microarrays [32], any difference would likely underestimate the strength of the cell-cycle oscillations (Figure 4). These results suggest that in humans, the relationship between $T_{rep}$ and S-phase transcription is substantially stronger than the well-established association with asynchronous expression.

**Discussion**

These results suggest that 1) S-phase transcription is associated with DNA replication timing in budding yeast, fission yeast, and human; 2) The association is strongest for genomic regions near ORIs, excluding the causal model in which $T_{rep}$ affects transcription [8-9, 25-26]; 3) It is also strongest for regions replicated in late S-phase, implying that early-firing ORIs are not affected by this relationship; and 4) This association explains at least three times more of the variability in $T_{rep}$ than the well-known association with (asynchronous) gene expression in human.

Although the replication of these patterns across three species (and across multiple data sets within species; Figures S1-S2) lends confidence to their robustness, several caveats should be considered. First, gene expression was represented by transcript abundances, which is a function of both transcription and mRNA decay; therefore the correlations reported here may underestimate the relationship between transcription and $T_{rep}$. This prediction can be tested once rates of transcription have been measured throughout the cell cycle. Second, data quality is critical in any analysis; poor-quality data can reduce, or entirely mask, a real relationship. However in most analyses reported here this is not a major concern, because it could only make the current results conservative (One exception to this is the ORI-proximal vs. distal analyses [Figures 3c and 4c]: if $T_{rep}$ was measured more accurately near ORIs, this would lead to stronger ORI-proximal correlations. Additional analysis suggests this is not the case [see Materials and Methods]). Third, correlation does not imply causation. Although the evidence does not support a model where $T_{rep}$ affects transcription (Figures 3c and 4c), I cannot determine whether transcription itself is affecting $T_{rep}$, or whether unobserved (latent) factors may be involved. With this caveat in mind, I believe there is still sufficient evidence to propose a testable model to account for these data.

A plausible mechanism explaining these observations draws from the finding that the firing of ORIs in late S-phase is governed by recruitment of limiting replication initiation factors [12,17-20]. These factors are sequestered by early-firing ORIs from G1 until early S-phase, and are reused at late-firing ORIs after their release from early-firing ORIs. I propose that the level of S-phase transcription near a late-firing ORI reflects local chromatin accessibility and/or subnuclear positioning, and in turn the ability of ORIs to recruit these limiting factors during S-phase (Figure 5). This model accounts for the relationship of $T_{rep}$ with S-phase transcription (and the differing relationships in other phases); for the relationship being strongest near late-firing ORIs; and for the inferred direction of causality (i.e. $T_{rep}$ not being causal).

The proposed mechanism likely acts in concert with other factors determining $T_{rep}$, and thus is not inconsistent with evidence supporting these other factors. For example, although the determination of early vs. late-firing ORIs is completed during M/G1 [12-14], S-phase transcription may still influence firing time specifically at late-firing ORIs (Figure 5). Future work integrating these results with other (non-mutually-exclusive) mechanisms affecting $T_{rep}$—e.g. Forkhead transcription factors [33] and subnuclear positioning [8,29,34-35]—may lead to a unified framework for understanding the causes, and consequences, of the temporal program of DNA replication across eukaryotes.

**Materials and Methods**

*Data sources*

Genome-wide $T_{rep}$ values were downloaded for all three species [2,4,6], and mapped onto genes by linear interpolation to the gene's midpoint. Asynchronous yeast expression levels (used in Figure 2) were taken from [36-37], using the poly-A data for *Sc* and the median of wildtype replicates for *Sp*. Asynchronous HeLa RNA-seq data were from the ENCODE project [31]. Identities of cell cycle-regulated genes, their expression levels, and the cell cycle phase of each expression time-point were acquired from [27-28,30]. All cell-cycle expression data were measured as mRNA levels relative to asynchronous levels of each gene, as opposed to absolute mRNA abundances that can be measured by RNA-seq; therefore these expression levels represent the relative induction or repression of each gene throughout the cell cycle. The order of maximum expression levels was obtained from [38] for *Sp* and [27] for *Sc*. ORI locations were downloaded from ORIdb [39] for both yeasts (using only "confirmed" or "likely" ORIs), and from [6] for human (see below).

*Yeast data analysis*

All correlations were Pearson's (significance cutoffs given in each figure legend). $T_{rep}$ moving averages (Figure 1c) were calculated for windows of 100 genes for *Sc* and 60 genes for *Sp* (due to the smaller number of cell cycle-regulated genes in *Sp*). For Figures 1a and 3a, the G2 expression data were represented by the 42 min time-point for *Sc* and 135 min for *Sp*; for Figure 1a, *Sc* M/G1 was represented by the 70 min time-point, and *Sp* G1 was represented by 225 min. For Figure 3b, the early/late S-phase cutoff was chosen at halfway through S-phase of each $T_{rep}$ data set (39.6 min after release from hydroxyurea arrest in *Sp*, and 26.8 min after release from *cdc7* arrest in *Sc*). The cutoff for ORI-proximal vs. ORI-distal (5 kb from each gene's 5' end in *Sc* and 10 kb in *Sp*) was chosen in each yeast to result in gene lists of approximately equal size.

P-values in Figure 2 were calculated with a two-tailed Student's t-test. Because the *Sc* expression levels were calculated as a ratio of mRNA/genomic DNA from asynchronous cells [37], it represents the number of mRNAs per DNA copy, and thus account for the fact that genes with early $T_{rep}$ spend a greater portion of the cell cycle with two copies. Although the *Sp* expression data [36] do not account for this, correcting for the effect by subtracting a fraction of each expression level proportional to the time each gene spends with two copies had only a minimal effect.

All code and data are available at http://www.stanford.edu/group/fraserlab/txn-rep.

*Human data analysis*

Human ORIs were defined as Orc1 binding sites [6] located within 1 mb of early-replicating peaks in the HeLa $T_{rep}$ profile, which indicate active ORIs (this window size was necessitated by the low resolution of the $T_{rep}$ profile) [6]. The early/late $T_{rep}$ cutoff was the first 50% of S-phase and the ORI-proximal/distal cutoff was 10 kb from each gene's 5' end. Due to the higher number of expression data points per cell cycle in human (~15 in human vs. ~9 for both yeasts), a two-point moving average was used for plotting human correlation coefficients.

To compare asynchronous expression vs. S-phase transcription in HeLa cells, I compared high-coverage RNA-seq data from HeLa cells [31] with $T_{rep}$ [6] for the same genes analyzed in Figure 4b-c. The fraction of variance in $T_{rep}$ explained by the expression data is simply the $r^2$ value from the Pearson's correlation. Comparing these values for the asynchronous data with the strongest G2-phase (used to represent S-phase transcription, as described above) correlations, among the late-replicating genes (represented by the red line in Figure 4b) 2.7% of the variance in $T_{rep}$ was explained by the asynchronous data, vs. 8.1% for S-phase transcription. Likewise for ORI-proximal genes (represented by the blue line in Figure 4c), the asynchronous data explained 2.3% of the variance in $T_{rep}$, vs. 7.6% for S-phase transcription.

To determine whether $T_{rep}$ is measured with greater accuracy near ORIs, I compared the $T_{rep}$ data used in Figure 4 [6] with an independent $T_{rep}$ data set from HeLa cells [40]. Restricting the analysis to the cell-cycle regulated genes analyzed in Figure 4c, I found that ORI-distal genes actually showed better agreement between $T_{rep}$ data sets than did ORI-proximal genes ($r = 0.59$ and 0.46, respectively). This implies that, if anything, $T_{rep}$ is measured less accurately in ORI-proximal regions, which would lead to an underestimate of the strength of the oscillating correlation (blue line in Figure 4c).

**Acknowledgements**

I would like to thank M. Botchan, A. Donaldson, D. Gilbert, M. Kobor, and J. Rine for helpful advice. This work was supported by NIH grant 1R01GM097171-01A1. HBF is an Alfred P. Sloan Fellow and a Pew Scholar in the Biomedical Sciences.

**Conflict of Interest**

The author declares no conflict of interest.

**Figure Legends**

**Figure 1. The transcription/$T_{rep}$ association varies by cell-cycle stage. a.** Comparing mean $T_{rep}$ of the top decile (10%) of most-induced vs. most-repressed cell cycle-regulated genes reveals that genes highly expressed in G2 replicate early in both *Sc* and *Sp*, whereas those highly expressed in M/G1 (*Sc*) or G1 (*Sp*) replicate late. **b.** The correlation between $T_{rep}$ and expression levels of known cell cycle-regulated genes was calculated, separately for expression levels from each time point of cell-cycle synchronized time courses [27-28]. An oscillation of the correlation coefficient (Pearson's *r*) was observed for both budding yeast (all $|r| > 0.107$ are significant at $p < 0.0025$) and fission yeast (all $|r| > 0.177$ are at $p < 0.0025$). The approximate cell-cycle phase of each time point is shown [27-28]. Similar oscillations are observed for other methods of synchronization as well (Figures S1-S2). **c.** A moving average of $T_{rep}$ is shown for all cell cycle-regulated genes, arranged in order of their time of maximal expression, beginning immediately following mitosis. A similar pattern is observed for both yeast species, with the latest $T_{rep}$ for genes with maximal transcript levels in G1, and the earliest $T_{rep}$ for genes with maximal transcript levels in G2.

**Figure 2. Asynchronous gene expression associates with $T_{rep}$ in budding and fission yeast.** Comparison of the 100 highest-expressed genes with the 100 lowest-expressed shows that highly expressed genes are replicated earlier in both budding yeast and fission yeast.

**Figure 3. Factors affecting the strength of the transcription/$T_{rep}$ association. a.** Median G2-phase transcript levels (representing S-phase transcription; Figure S4a) are shown for all cell cycle-regulated genes separated into ten equally-sized bins (deciles) by their $T_{rep}$. For both yeast species, no correlation is observed for the first five bins, whereas as strong relationship is present for later

$T_{rep}$. **b.** Consistent with the decile analysis, no oscillation is observed in the correlation between expression level and $T_{rep}$ for early-$T_{rep}$ genes, while a strong oscillation is observed for late-$T_{rep}$ genes. **c.** Only weak oscillation is observed in the correlation between expression level and $T_{rep}$ for ORI-distal genes (greater than 5 kb from the nearest ORI in budding yeast, or 10 kb in fission yeast), while a strong oscillation is observed for ORI-proximal genes.

**Figure 4. Transcription and $T_{rep}$ in human. a.** No oscillation is observed when comparing the $T_{rep}$ vs. expression levels of all cell-cycle regulated genes in HeLa cells (all $|r| > 0.063$ are significant at $p < 0.05$; the four time points that exceed this are within the range expected by chance, given that 47 time points were analyzed). **b.** Significant oscillation is observed when comparing the $T_{rep}$ vs. expression levels of cell-cycle regulated genes with late $T_{rep}$ (red line; the final 50% of S-phase; all $|r| > 0.195$ are significant), but not early $T_{rep}$ (blue line). **c.** Significant oscillation is observed when comparing the $T_{rep}$ vs. expression levels of cell-cycle regulated genes within 10 kb of an ORI (blue line; all $|r| > 0.197$ are significant), but not further than 10 kb from an ORI (red line).

**Figure 5. A model to explain these observations. Components:** ORC and MCM2-7 are protein complexes comprising the pre-replicative complex. Blue cylinders represent nucleosomes, with dark blue indicating closed/repressive chromatin and light blue indicating open/accessible chromatin. Red proteins are limiting replication initiation factors (such as Cdc45 and Sld3). Txn = transcription. **Sequence of events:** In G1 (not depicted), the limiting replication initiation factors (red circles) associate with the earliest-firing ORIs (top row). When S-phase begins, these early ORIs fire and release the factors, which are then free to associate with other ORIs (though note that Cdc45 is a component of the replication fork, so can only be recycled after fork termination). The relative affinities of the remaining ORIs for these factors—and thus their relative firing times—are determined by the chromatin state near the ORI during S-phase. ORIs near genes highly transcribed in S-phase (middle row) have an accessible chromatin structure and thus high affinity, so will tend to fire earlier than those with little nearby S-phase transcription and thus less accessible chromatin (bottom row). Although not shown here, subnuclear positioning could help determine ORI accessibility, either by influencing chromatin structure or through other mechanisms. Figure adapted from [19].

**Figure S1.** Correlation analysis as in Figure 1b (left), but using a different method for synchronization of *Sc* (a temperature-sensitive *cdc28* mutant).

**Figure S2.** Correlation analysis as in Figure 1b (right), but using a different method for synchronization of *Sp* (a temperature-sensitive *cdc25* mutant).

**Figure S3.** Repeating the moving average analysis from Figure 1c, with standard errors shown for each point (grey lines) in **a.** *Sc* and **b.** *Sp*. Results suggest the differences between high and low windows are unlikely to be due to random fluctuations.

**Figure S4. a.** The mean expression level of 100 genes comprising the window with earliest $T_{rep}$ in Figure 1c (left) is plotted as a function of time in the cell cycle. The genes that reach a maximum mRNA level in G2 have their maximum rate of increase (and likely maximum rate of transcription) in S phase. **b.** As for part (a), but showing the mean expression for the 100 genes in the window with the latest $T_{rep}$ in Figure 1c (left). The genes that reach a minimum mRNA level in G2 have their maximum rate of decrease (and likely minimum rate of transcription) in S phase.

**Figure S5.** Repeating the ORI proximal/distal analysis from Figure 3c using a cutoff of 7.5 kb to define ORI proximal in **a.** *Sc* and **b.** *Sp.* Results are qualitatively identical to Figure 3c.

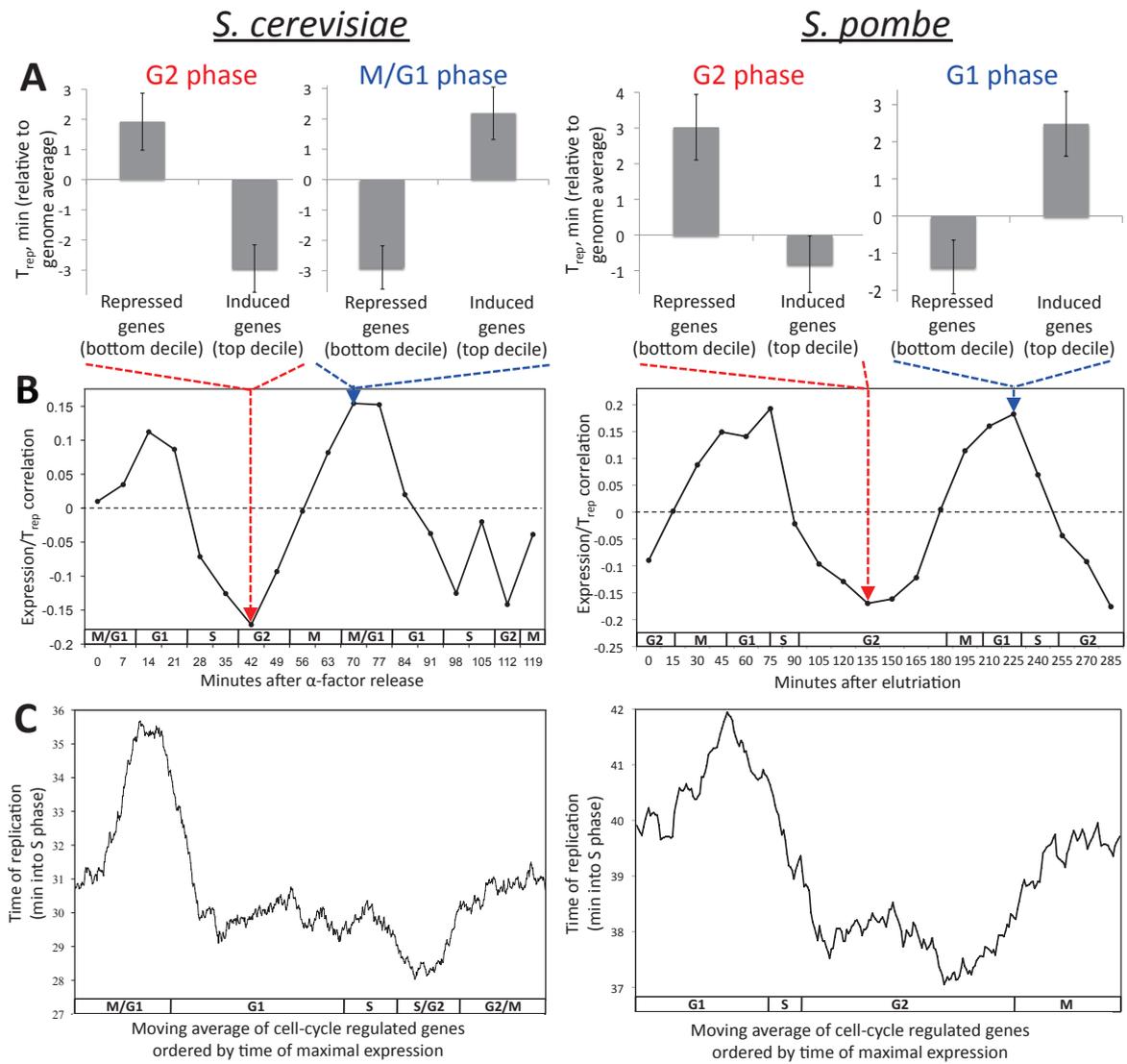

Figure 1

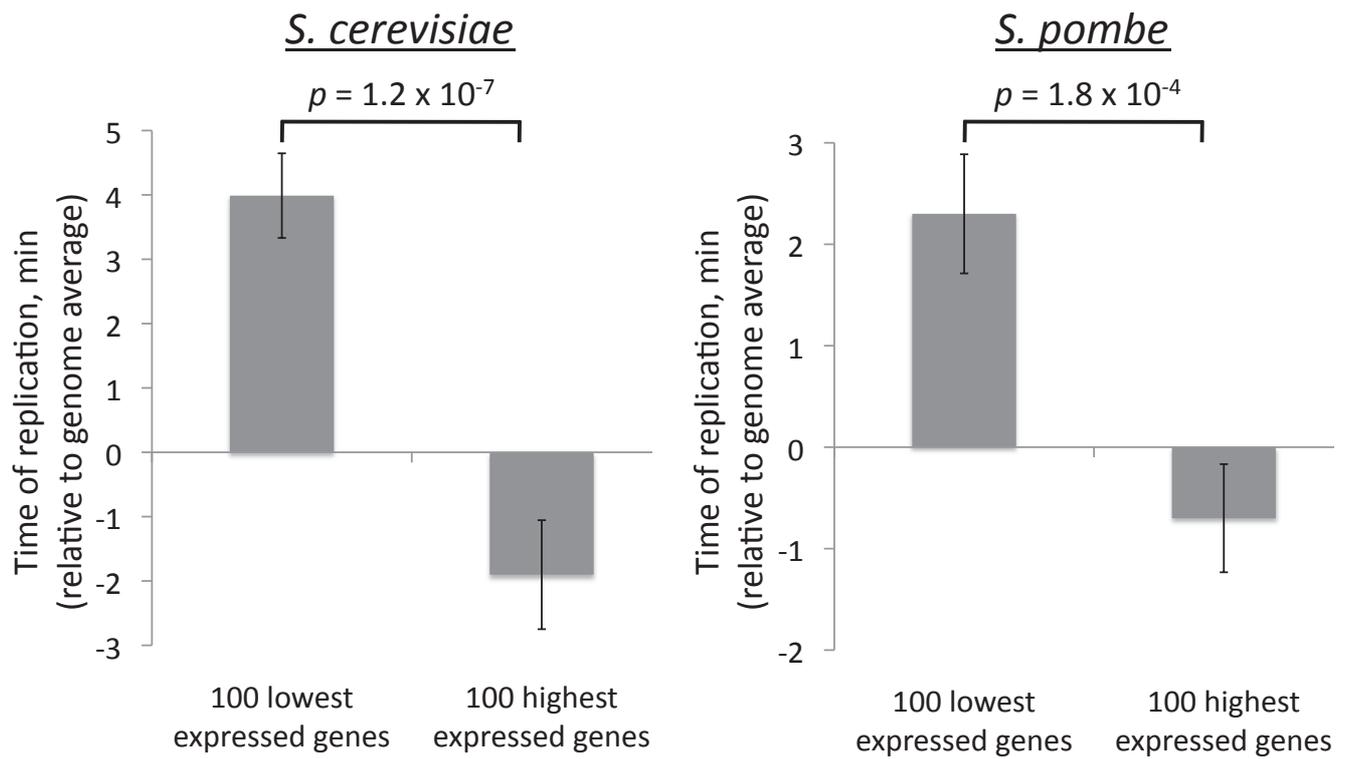

Figure 2

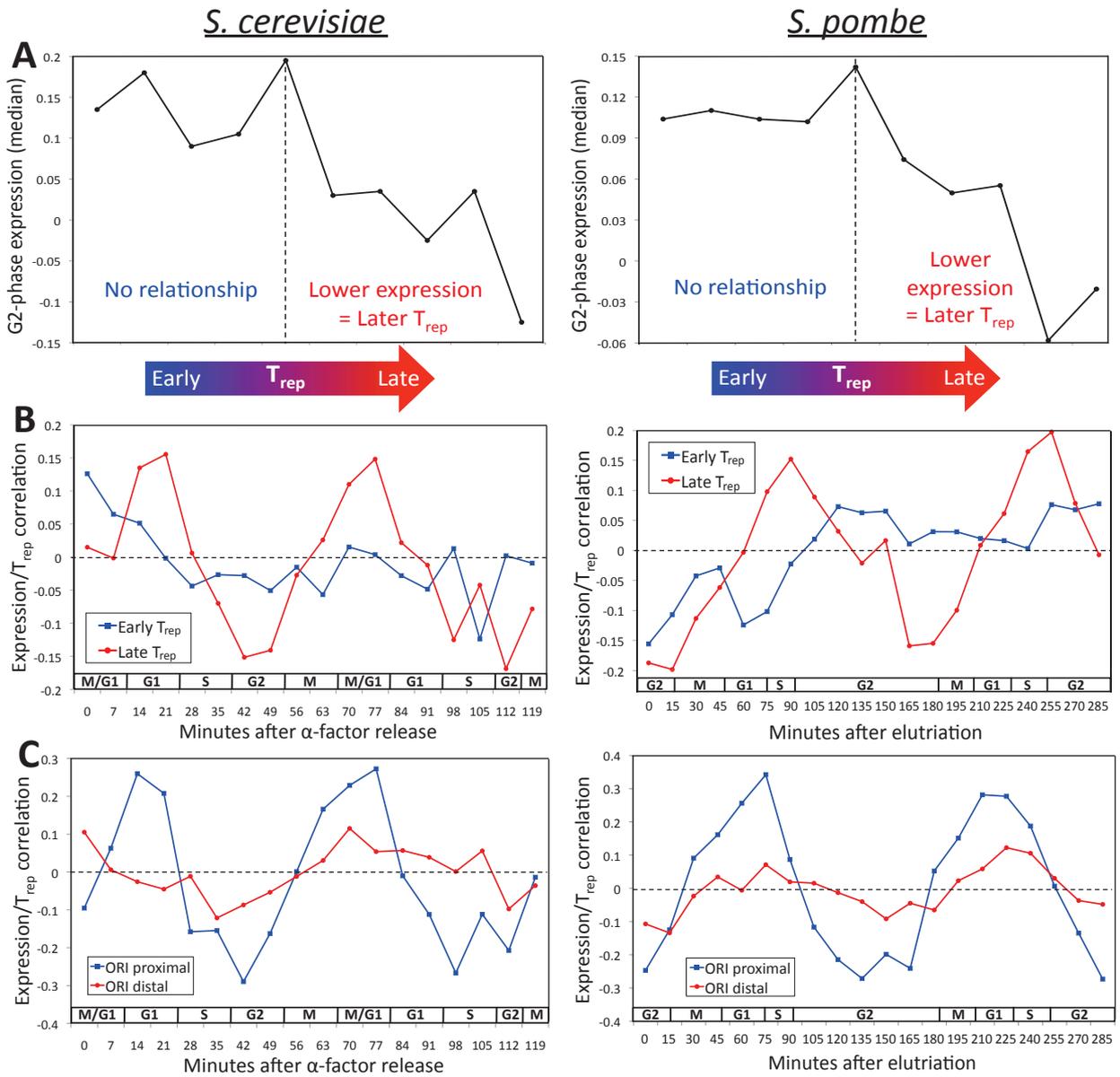

Figure 3

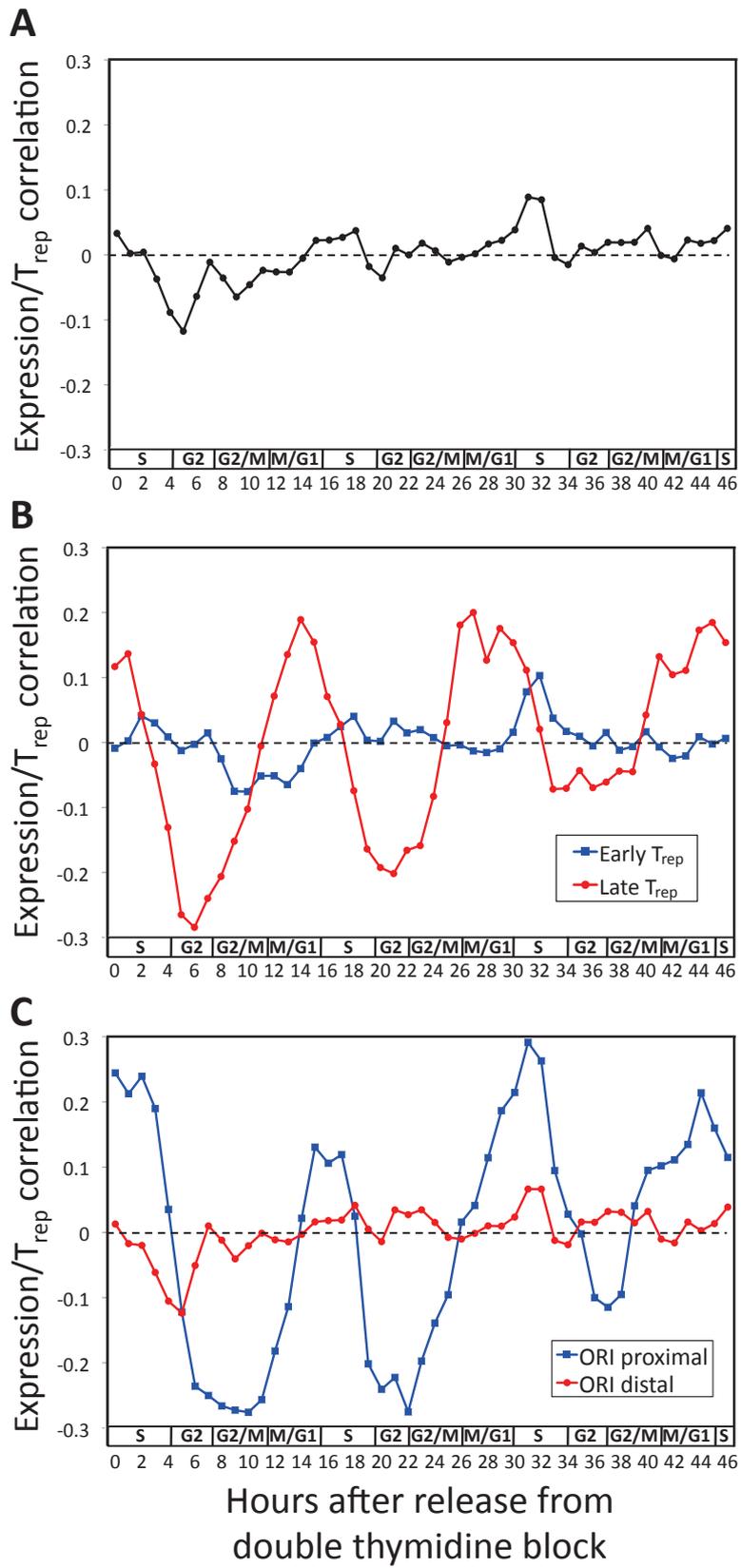

Figure 4

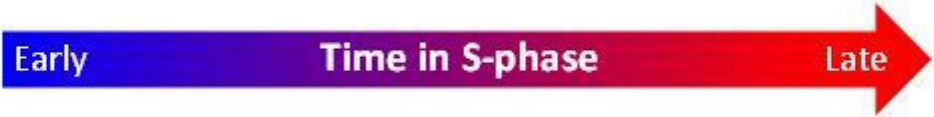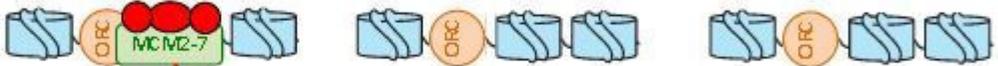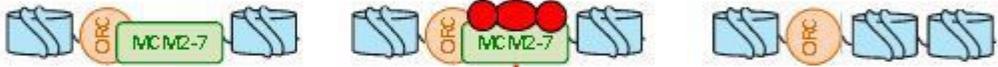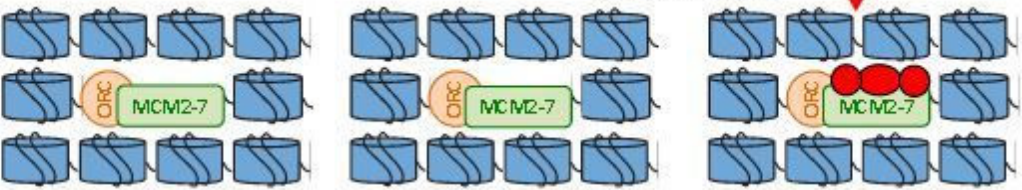

Figure 5

**Additional files provided with this submission:**

Additional file 1: txn-rep supp figs.pdf, 207K
http://genomebiology.com/imedia/9433564991052221/supp1.pdf